\documentclass{PoS}

\usepackage{graphicx}
\usepackage{mathrsfs}
\usepackage{amsmath}

\title{On the Analytic Structure of Scalar Glueball Operators}

\ShortTitle{On the Analytic Structure of Scalar Glueball Operators}

\author{\speaker{Andreas Windisch}\\
        Institut f\"ur Physik, Karl-Franzens Universit\"at Graz, Universit\"atsplatz 5, 8010 Graz, Austria\\
        E-mail: \email{andreas.windisch@uni-graz.at}}

\author{Markus Q. Huber\\
        Institut f\"ur Kernphysik, Technische Universit\"at Darmstadt, Schlossgartenstrasse 2, 64289 Darmstadt, Germany\\
        E-mail: \email{markus.huber@physik.tu-darmstadt.de}}

\author{Reinhard Alkofer\\
        Institut f\"ur Physik, Karl-Franzens Universit\"at Graz, Universit\"atsplatz 5, 8010 Graz, Austria\\
        E-mail: \email{reinhard.alkofer@uni-graz.at}}

\abstract{The correlator of the square of the Yang-Mills field-strength tensor corresponds to a scalar glueball, i.~e., to a bound-state formed by gluonic ingredients only.
It has quantum numbers $0^{++}$ and its mass, as predicted by different theoretical approaches, is expected to lie between 1 and 2 GeV. Here we restrict our considerations to the Born level, that is, we consider the correlator to zeroth order in the coupling.
Gluonic self-interaction is taken into account indirectly by using non-perturbative gluon propagators. The employed closed expressions are motivated by lattice and Dyson-Schwinger studies.
The analytic continuation of the integrals themselves is complicated by additional obstructive structures like branch cuts and poles that are induced by the inner integral in the complex plane of the outer integration variable. We deal with this problem by deforming the outer integration contour accordingly. For different input gluon propagators we find a positive glueball spectral density which is required for physical states. Poles are, however, absent which is most likely an artifact of working at Born level.}

\FullConference{Xth Quark Confinement and the Hadron Spectrum,\\
		October 8-12, 2012\\
		TUM Campus Garching, Munich, Germany}

\begin{document}
\section{\label{intro}Introduction}

As color-carrying states gluons do not appear as asymptotic physical states. They are confined to observable color singlet objects by some mechanism, see, e.~g., \cite{Alkofer:2006fu} for a short review. In pure Yang-Mills theory the only possibility to generate color neutral -- and thus observable -- states is to combine several gluons to form a bound state, a glueball. Experimentally, they are very hard to find due to mixing with mesonic states. On the theoretical side, several approaches for gluonic bound states are available, see, for instance, \cite{Mathieu:2008me} for a recent review.\par  

In the following we calculate the correlator of a $0^{++}$ glueball candidate and
determine its analytic properties \cite{Windisch:2012sz} from which we can extract the spectral density. 
For a physically observable state the spectral density must be positive in order to allow a probabilistic interpretation \cite{Kallen:1952zz,Lehmann:1954xi}. Thus, the detection of positivity violations indicates that a certain state is expelled form the asymptotic state space (and in this sense confined), while the converse does not hold necessarily. Glueballs must therefore possess a positive spectral density, although their constituents may not. Indeed, positivity violations of gluons are established in the Landau gauge from lattice \cite{Langfeld:2001cz,Bowman:2007du} and functional calculations \cite{Alkofer:2003jj,Fischer:2008uz}.

As a first approximation we take into account only the Abelian part of the field strength tensor. This simplifies the calculations in several respects. For example, renormalization would become more complicated but the required machinery is available \cite{Dudal:2008tg}. However, the input we use was obtained from full Yang-Mills theory and therefore contains interactions. In the following we will use two different fits for the gluon propagator and calculate the glueball correlator numerically. In simple cases this is also possible analytically, see, e.~g. \cite{Zwanziger:1989mf,Baulieu:2009ha}. For future applications a numeric procedure is certainly advantageous as it allows, for instance, to use also numerical data as became available only recently \cite{Strauss:2012dg}.

The calculation of the correlator boils down to a two dimensional integral. Since we consider complex external momenta, an additional subtlety arises: The inner integral leads to non-analytic structures, like branch cuts, in the complex plane of the outer integration variable which have to be taken into account properly. We do this here by deforming the integration contour, i.~e., also the radial integration variable becomes complex. This method is already known, see, for example, \cite{Alkofer:2003jj}, but in the present case the structure of the arising non-analyticities is especially tedious as detailed below.

The two gluon propagator fits we employ here are of the decoupling
\cite{Dudal:2007cw,Boucaud:2008ji,Aguilar:2008xm,Fischer:2008uz,Alkofer:2008jy} and
scaling type \cite{vonSmekal:1997is}. For the former we use a fit to lattice data
motivated by the refined Gribov-Zwanziger scenario (RGZ) \cite{Cucchieri:2011ig}. For
the chosen parameter values it possesses two complex conjugate poles. Note that other
fits, for instance, in \cite{Aguilar:2007ie}, were suggested as well, but most of them
are special cases of the used one. For the scaling type propagator we use a fit to the solution of a Dyson-Schwinger study \cite{Alkofer:2003jj}. It has a branch cut on the negative real axis. 
\section{\label{prerequisites}Some Prerequisites}
We consider the correlator of a candidate for a scalar glueball with quantum numbers $0^{++}$, 
\begin{equation}
\langle F^2(x)F^2(0)\rangle_d = \langle F_{\mu\nu}^a(x)F_{\mu\nu}^a(x)F_{\rho\sigma}^b(0)F_{\rho\sigma}^b(0)\rangle_d,
\label{eq1}
\end{equation}
where $d$ is the space-time dimension and $F_{\mu\nu}^a(x)$ is, as part of our approximation, just the Abelian part of the Yang-Mills field-strength tensor given by
\begin{equation}
F_{\mu\nu}^a=\partial_\mu A^a_\nu-\partial_\nu A^a_\mu.
\label{eq2}
\end{equation} 
We are interested in the momentum space representation of this correlator,
\begin{equation}
\langle F^2(x)F^2(0)\rangle_d=\int\frac{d^dp}{(2\pi)^d}e^{i\,p\cdot x}\mathscr{O}_d(p^2).
\label{eq3}
\end{equation}
The desired expression, $\mathscr{O}_d(p^2)$, reads \cite{Baulieu:2009ha}
\begin{equation}
\mathscr{O}_d(p^2)=8(N_C^2-1)\int \frac{d^dk}{(2\pi)^d}\left(\mathscr{G}((p-k)^2)\mathscr{G}(k^2)(k^2(p-k)^2+(d-2)(k\cdot(p-k))^2)\right).
\label{eq4}
\end{equation} 
For a transverse gluon propagator we have
\begin{equation}
D_{\mu\nu}(p^2)=\left(\delta_{\mu\nu}-\frac{p_\mu p_\nu}{p^2}\right)\mathscr{G}(p^2),
\label{eq5}
\end{equation} 
 where only the scalar part $\mathscr{G}(p^2)$ enters the expression (\ref{eq4}). A further complication we have not addressed so far is the fact that in 4 Euclidean space time dimensions, the integral as given in eq.~(\ref{eq4}) diverges like $\sim p^4$. To render the integral finite we employ the BPHZ renormalization, i.~e.,  we Taylor subtract the divergent terms:
\begin{equation}
 \mathscr{O}^r_d(p^2)=\mathscr{O}_d(p^2)-\mathscr{O}_d(0)-p^2\frac{\partial^2}{\partial p^2}\mathscr{O}_d(p^2)\Big|_{p=0}-p^4 \frac{\partial^4}{\partial p^4}\mathscr{O}_d(p^2)\Big|_{p=0}.
\label{eq6}
\end{equation}
The odd derivatives vanish because of the anti-symmetry of the angular integral. In order to obtain the analytic structure of the scalar glueball correlator, we have to solve eq.~(\ref{eq6}) for complex values of the square of the external momentum. The spectral density is then accessible by evaluating the discontinuity of the branch cut along the negative real axis. For the two-point function $\Delta(p^2)$ of a given spin zero operator $\Phi$, the spectral density reads
\begin{equation}
 \rho(p^2)=\frac{1}{2\,\pi\,i}\lim_{\epsilon\rightarrow 0^+}[\Delta(-p^2-i\,\epsilon)-\Delta(-p^2+i\,\epsilon)]
\label{eq7}
\end{equation}
and the spectral representation of the two-point function is
\begin{equation}
 \Delta(p^2)=\int\frac{d^dp}{(2\pi)^d}e^{i\,p\cdot x} \langle \Phi(x) \Phi(0) \rangle=\int_{\tau_0}^\infty d\tau\frac{\rho(\tau)}{\tau+z},
\label{eq8}
\end{equation}
if no poles or cuts except for time-like momenta exist. $\tau_0$ is the multi-particle threshold.

Eq.~(\ref{eq4}) holds for arbitrary dimensions. Here we consider only $d=4$. The two-dimensional case, which of course has a trivial glueball spectrum, served as a test-case for the development of the numerics and is presented together with the four-dimensional results in \cite{Windisch:2012sz}.
\section{\label{method}The Method}
The algorithm we use here is described in detail in \cite{Windisch:2012zd}, where as an example the analytical results from \cite{Baulieu:2009ha} were reproduced. Let us consider the case of the RGZ propagator fit of \cite{Cucchieri:2011ig},
\begin{equation}
\mathscr{G}(p^2)=C\frac{p^2+s}{p^4+u^2p^2+t^2}.
\label{eq9}
\end{equation}
The fit-parameters are $s=2.508\,\mathrm{GeV^2}$, $t=0.72\,\mathrm{GeV^2}$, $u=0.768\,\mathrm{GeV}$ and $C=0.784$ \cite{Cucchieri:2011ig}. In \cite{Windisch:2012zd} the following steps are given in order to evaluate the integral (\ref{eq4}):
\begin{itemize}
\item STEP 1: \textit{Express (\ref{eq4}) in hyper-spherical coordinates}\\
\begin{eqnarray}
\label{eq10}
\mathscr{O}_4(x)=\frac{8C^2}{\pi^3}\int_0^\infty &dy&\ y\int_{-1}^1dz\sqrt{1-z^2}\frac{x+y-2\sqrt{x}\sqrt{y}\,z+s}{(x+y-2\sqrt{x}\sqrt{y}\,z)^2+u^2(x+y-2\sqrt{x}\sqrt{y}\,z)+t^2}\nonumber\\
&\times&\frac{y}{y^2+u^2y+t^2}\left[(x+y-2\sqrt{x}\sqrt{y}\,z)y+2(\sqrt{y}\sqrt{x}\,z-y)^2\right],
\end{eqnarray}
where $x=p^2$, $y=k^2$ and $p\cdot k=\sqrt{x}\sqrt{y}\,z$.
\item STEP 2: \textit{Renormalization}\\
The integral (\ref{eq10}) diverges quadratically in $x$. The renormalized expression is given by (\ref{eq6}). 
\item STEP 3: \textit{Analytic continuation}\\
For the present case this step can be performed either analytically or numerically. For $x\in\mathbb{C}$ the inner integral of eq.~(\ref{eq10}) can produce an integrable singularity together with the rest of the integrand. When $z$ runs through its integration interval ${[-1,1]}$, it picks up a whole line of these singular points resulting in a branch cut in the complex plane of the radial integration variable $y$. Thus the contour of the radial integral has to be deformed in order to avoid the cut. For eq.~(\ref{eq10}) we find two branch cuts as well as a pair of complex conjugate poles. The branch cuts, parametrized by $z$, in the $y$-plane can be determined analytically by finding the zeroes of the integrand of eq.~(\ref{eq10}) for a given $x\in\mathbb{C}$. We compared these results with a numerical integration. For $x=-2+2i$ both are shown in Fig.~\ref{fig1}.
\begin{figure}[tb]
\centering
\includegraphics[width=4.5cm]{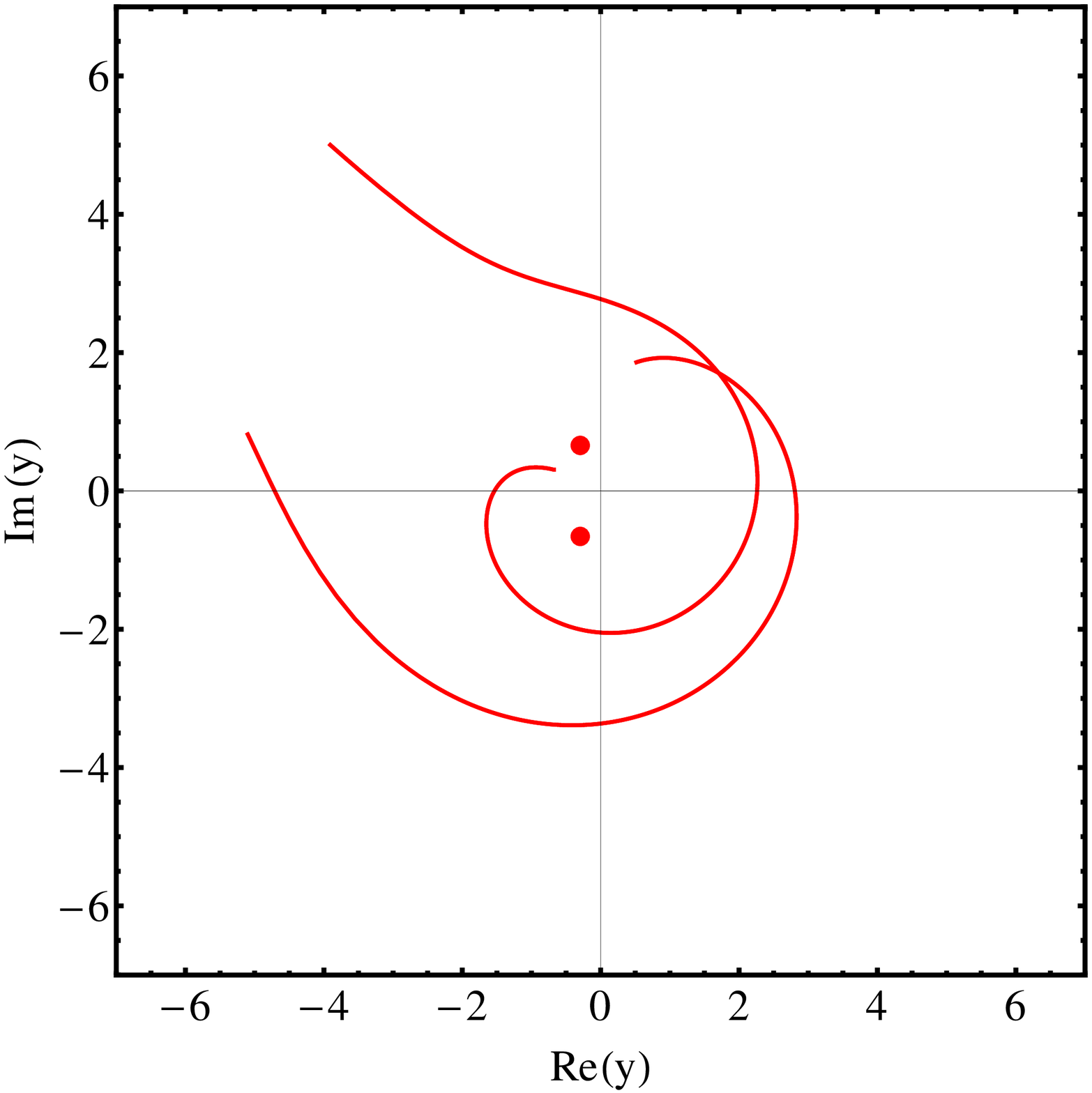}\hspace{1cm}
\includegraphics[width=5cm]{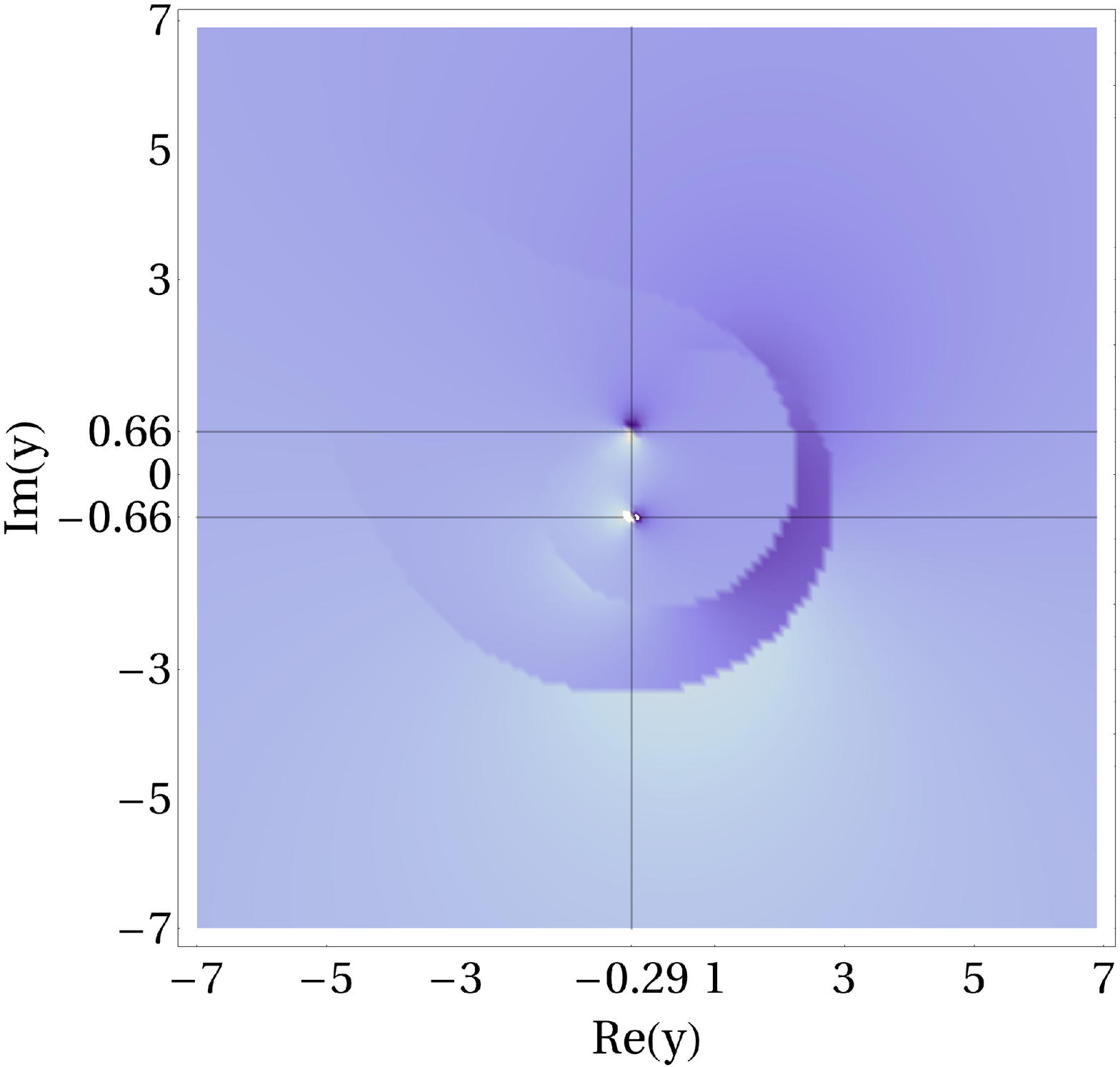}
\caption{\textit{Left:} Analytic results for the branch cuts and poles in the complex $y$-plane for $x=-2+2i$. \textit{Right:} Numerical verification of the analytic result.} 
\label{fig1}
\end{figure}

It is clearly visible in Fig.~\ref{fig1} that the deformation of the contour of the $y$-integration, required to connect $y=0$ to $y=\xi^2$ where $\xi$ is a UV cutoff, can be quite tricky. In general the open piece between the branch cuts always points in the direction of $\mbox{Arg}(x)$. Thus, if $x$ is on the positive real axis, the integration is straightforward since $y$ can be kept real as well. Now let us consider a complex $x=(r,\phi)$ by keeping $r$ fixed while $0<\phi<\pi/2$. There are no poles in the first quadrant, and the opening of the branch cuts always point in the direction of $\mbox{Arg}(x)$, thus the contour can be deformed continuously in that case. The same is true for the fourth quadrant. However, the complex conjugate poles of the integrand located in the quadrants II and III require more care. Obviously the contour cannot be deformed as easily for $\mbox{Arg}(p_{III})>\mbox{Arg}(x)>\mbox{Arg}(p_{II})$, where $p_{II}$ and $p_{III}$ are the pole locations in the second and third quadrants, respectively. For some values of $x$ the branch cut end points are narrowing down the area for a possible contour, see Fig.~\ref{fig2}. When an endpoint of a branch cut coincides with one of the poles, the contour cannot be deformed continuously and a non-analyticity arises in the integral. In \cite{Windisch:2012sz} we confirmed that for all points where this happens a branch point is also predicted from the Cutkosky rules \cite{Dudal:2010wn}.    
\begin{figure}[tb]
\centering
\includegraphics[width=4.5cm]{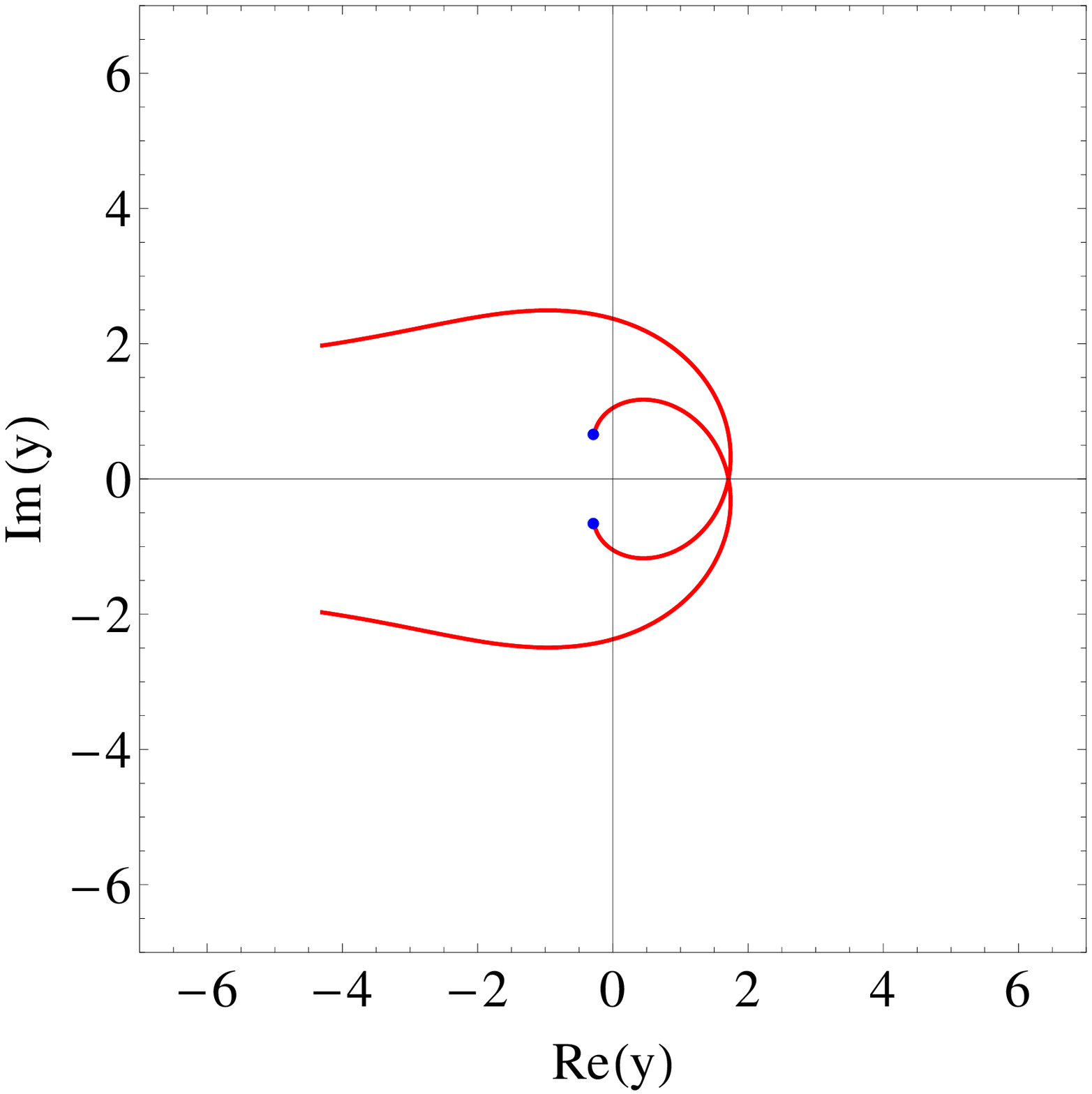}\hspace{1cm}
\includegraphics[width=4.5cm]{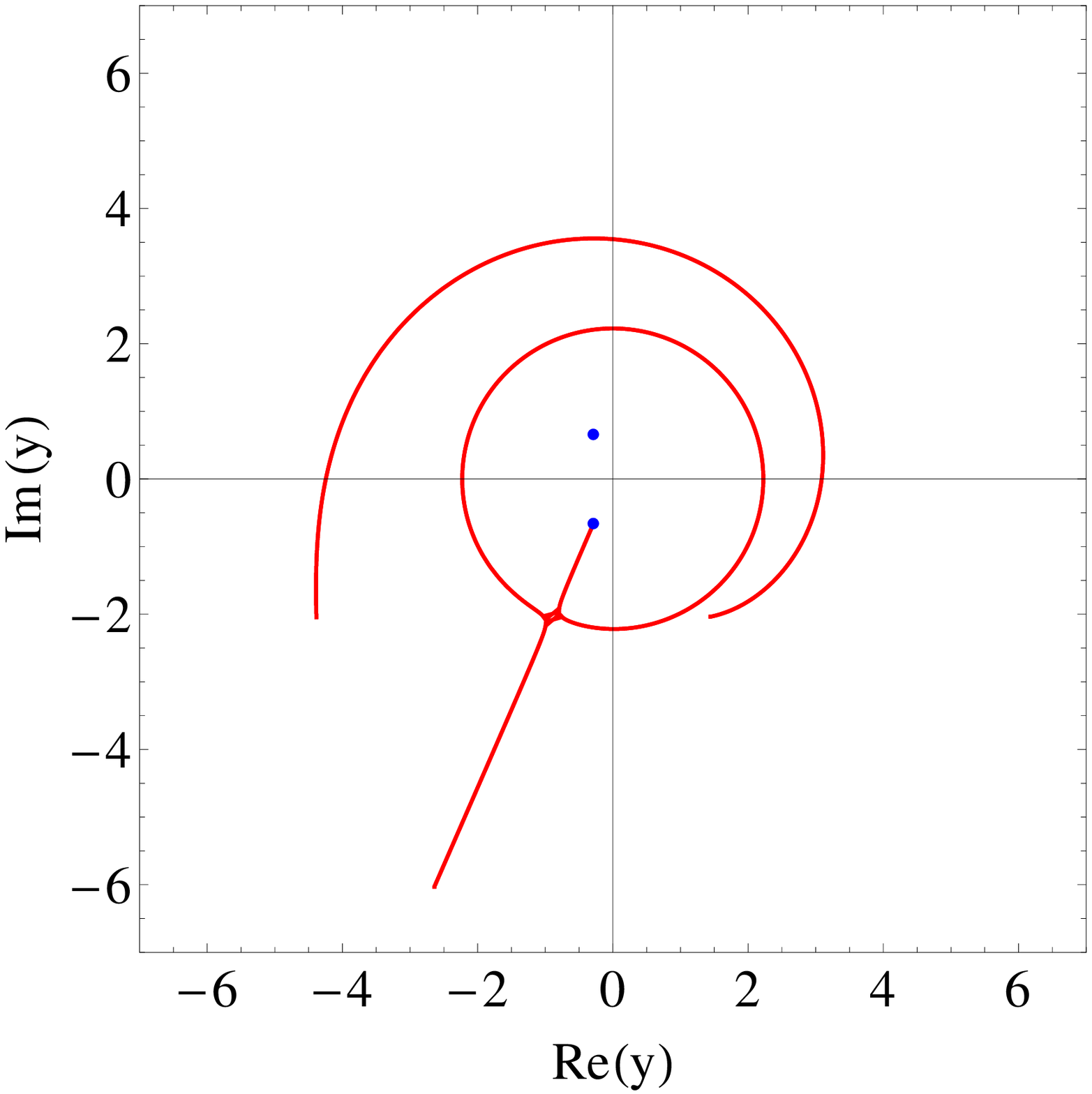}
\caption{\textit{Left:} The region for possible contours narrows down, $x=-2$. \textit{Right:} $x=-1.18+2.7i$. The same situation occurs for $x=-1.18-2.7i$, with the cuts flipped around the real axis.} 
\label{fig2}
\end{figure}
\end{itemize}
There are two further steps which we omit here as they are purely technical. What is relevant here is that the complex conjugate poles together with the two branch cuts severely restrict the possibilities for the contour deformation. It is hard to obtain stable results for complex values of $x$ when the argument of $x$ coincides with the argument of one of the pole locations. As discussed in the next section, we find three branch cuts for the RGZ case, one along the negative real axis, and two along the directions $\mbox{Arg}(p_{II})$ and $\mbox{Arg}(p_{III})$. The numerical determination of the branch points in this case is very troublesome, because for $x$-values close to the cuts in the $x$-plane the contour necessarily always comes very close to the cuts in the $y$-plane what leads to numerical artifacts. Even though the scaling propagator of \cite{Alkofer:2003jj} has a branch cut and the integrand induces two more cuts in the $y$-plane, the absence of poles allows a continuous contour deformation to values very close to the negative real axis. The results for the scaling propagator are thus not plagued by numerical issues. 
\section{\label{results}Results}
\subsection{\label{RGZ}Decoupling}
\begin{figure}[tb]
\centering
\includegraphics[width=7cm]{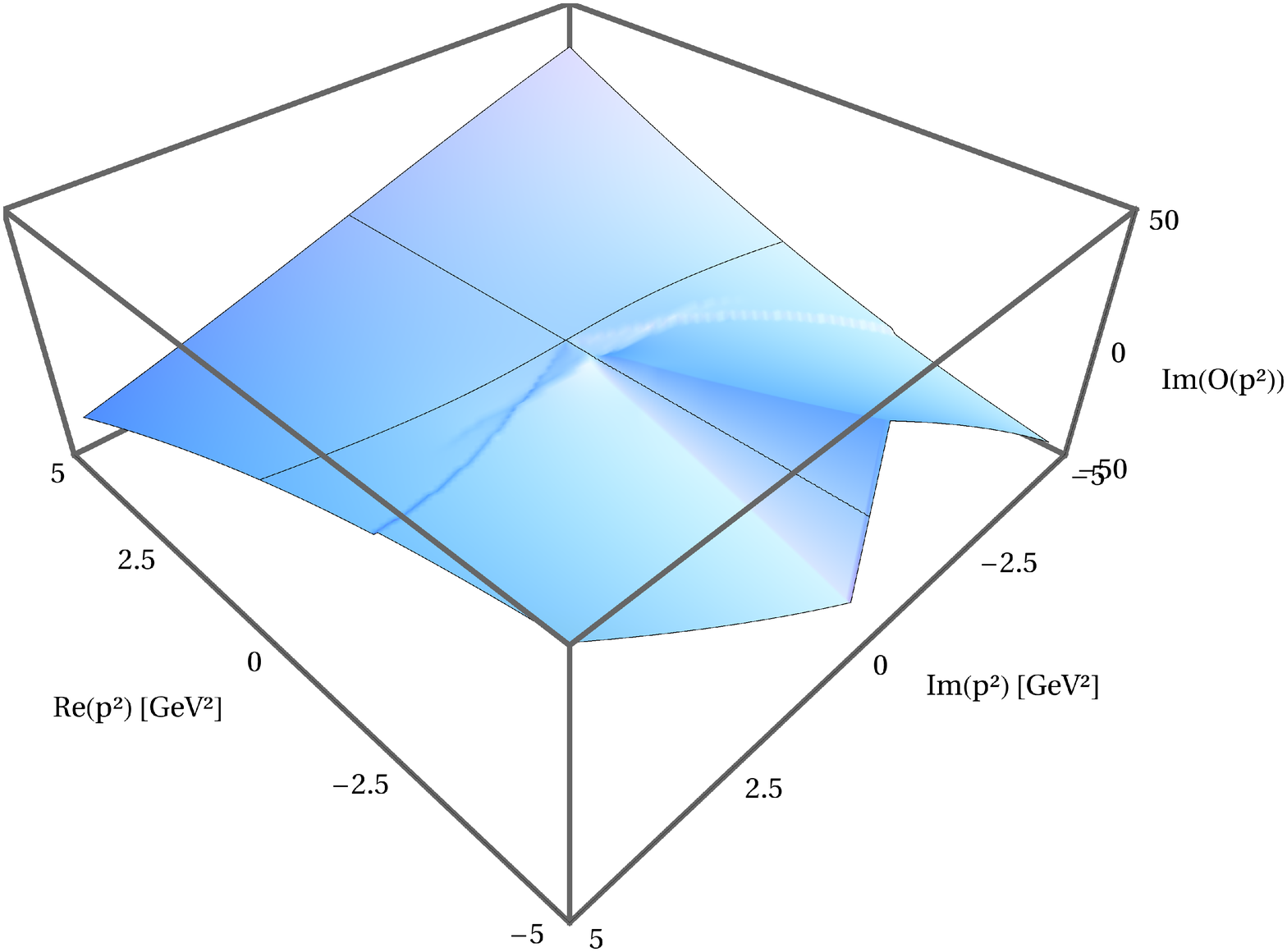}\hspace{1cm}
\includegraphics[width=7cm]{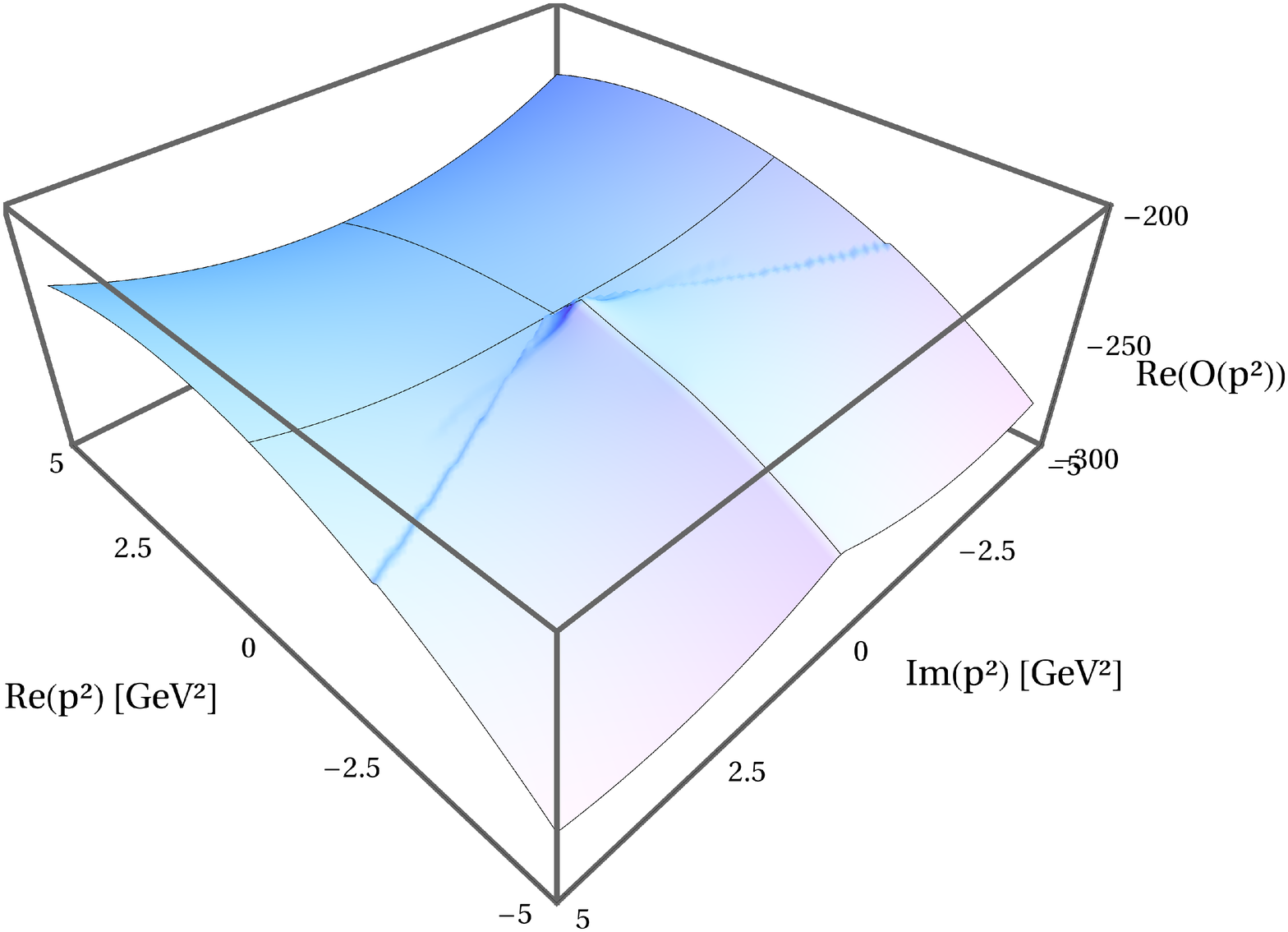}
\caption{\textit{Left:} The imaginary part of the scalar glueball correlator with RGZ gluons as input. \textit{Right:} The real part of the correlator.} 
\label{fig3}
\end{figure}

\begin{figure}[b]
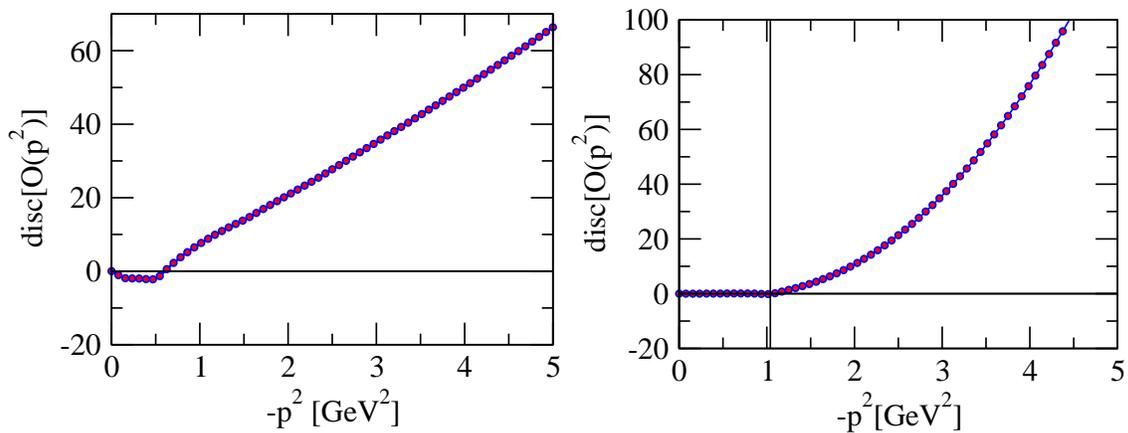

\begin{center}
\begin{minipage}{0.49\textwidth}
\includegraphics[width=\textwidth]{disc_RGZ_4d.eps}
\end{minipage}
\begin{minipage}{0.49\textwidth}
\includegraphics[width=\textwidth]{disc_prd70_4d.eps}
\end{minipage}
\caption{\label{fig4}The discontinuity of the physical branch cuts. \textit{Left:} Decoupling gluons. \textit{Right:} Scaling gluons.}
\end{center}
\end{figure}

In the previous section we already discussed several aspects of the RGZ propagator as gluonic input. Most importantly, we confirmed the location of the branch points known from the Cutkosky rules. 
Fig.~\ref{fig3} shows the imaginary and real parts of the correlator. The three branch cuts are clearly visible. The two 'unphysical` ones open very slowly. The extracted discontinuity of the 'physical' branch cut is depicted in fig.~\ref{fig4}. It becomes negative for small values of $-p^2$ and rises earlier than expect from the Cutkosky analysis. From investigating the complex plane of the radial integration variable we know that these phenomena are numerical artifacts which we expect to vanish if the contour deformation is better tuned; see \cite{Windisch:2012sz} for a more detailed discussion. Thus we conclude that the spectral density is positive.


\subsection{\label{scaling}Scaling}
The IR part of the scaling gluon fit is given by \cite{Alkofer:2003jj},
\begin{align}\label{eq:prop-scal}
 \mathscr{G}(p^2)=w\frac{1}{p^2}\left(\frac{p^2}{p^2+\Lambda^2}\right)^{2\kappa},
\end{align}
with $\kappa=0.595353$. The exponent leads to a branch cut of the propagator for complex momenta. We neglected the UV part of the propagator fit, which involves a logarithm, as we are only interested in IR relevant parts of the propagator. The other parameters are $w=2.5$ and $\Lambda=0.51$~GeV.
\begin{figure}[tb]
\centering
\includegraphics[width=7cm]{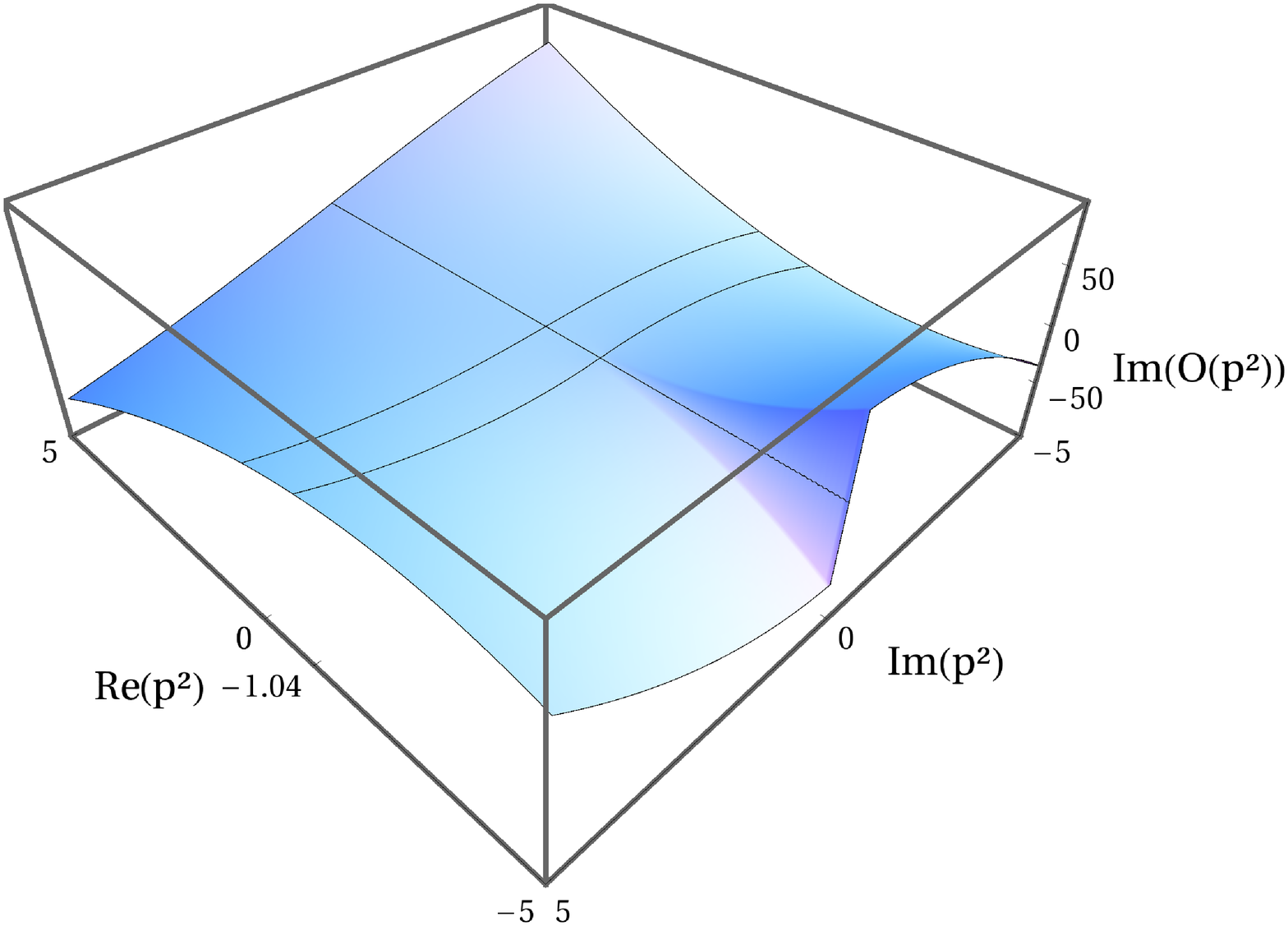}\hspace{1cm}
\includegraphics[width=7cm]{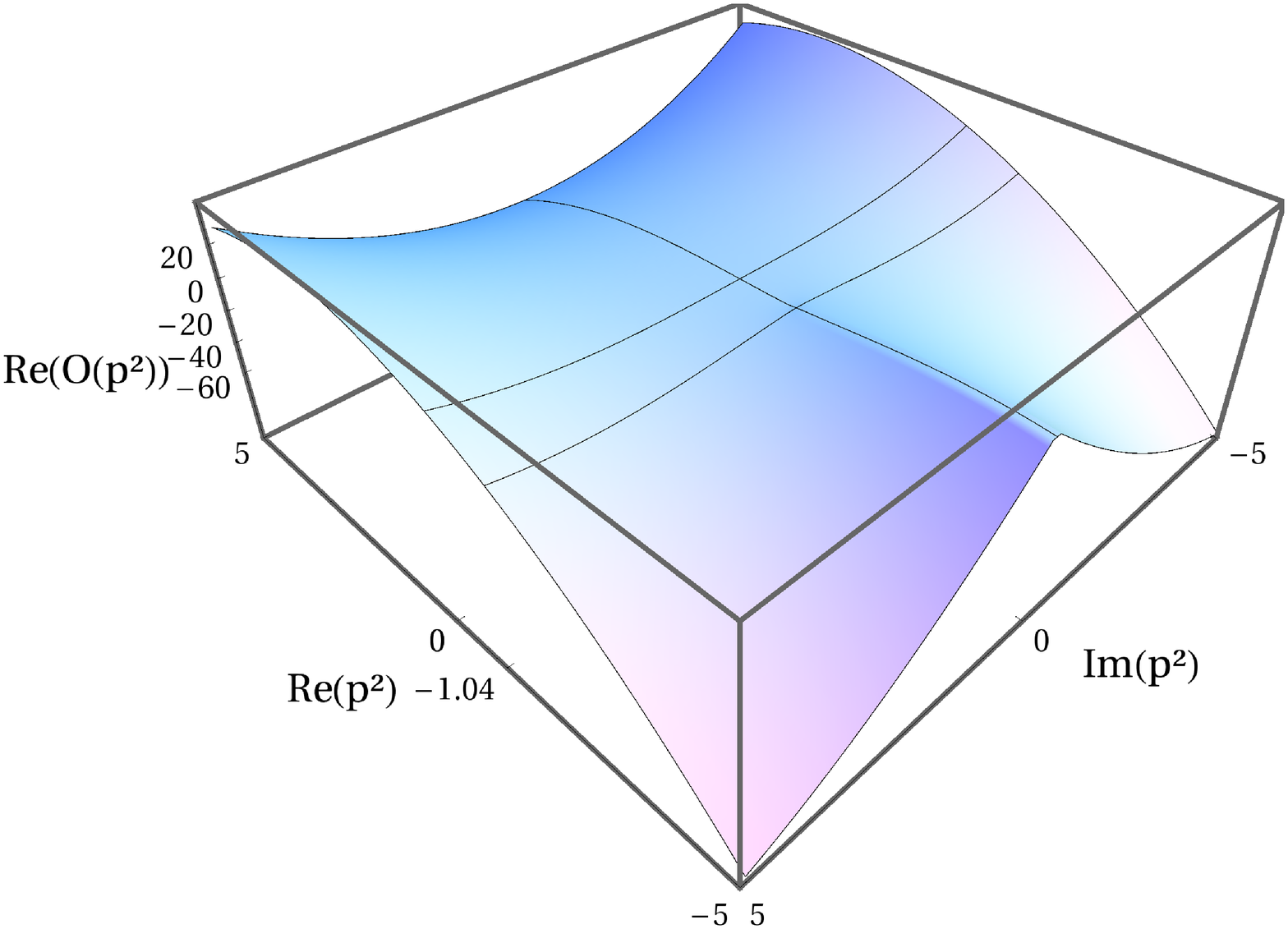}
\caption{\textit{Left:} The imaginary part of the scalar glueball correlator with scaling gluons as input. \textit{Right:} The real part of the correlator.} 
\label{fig5}
\end{figure}

Fig.~\ref{fig5} shows the imaginary and the real parts of the correlator. Since there are no non-analyticities besides the branch cut on the negative real axis, a K\"all\'en-Lehmann representation is possible. The corresponding positive spectral density is depicted in fig.~\ref{fig4}. We also observe that in this case the evaluation of the correlator in the complex plane is not plagued by numerical artifacts.
Strictly speaking the Cutkosky analysis is in this case not applicable, since the employed propagator does not have the required form. However, a naive application leads to a threshold in precise agreement with our numeric result.

\section{\label{summary}Summary}
In this work we studied the analytic properties of a scalar glueball correlator at Born-level. The self-interaction of gluons entered via using non-perturbative gluon propagator fits. These exhibit positivity violations and describe thus confined gluons. The resulting glueball correlators have a branch cut for time-like momenta and no poles. The extracted spectral densities are positive as required for a physical state. For the fit of the decoupling propagator we also find two unphysical cuts which are due to the analytic structure of the fit. Possible continuations include the addition of higher order terms and the use of numerical results for the propagator in the complex plane. The employed techniques for the evaluation of the integrals may be useful for other studies, where complex momenta are involved, as well.

\acknowledgments
MQH acknowledges support by the Alexander von Humboldt foundation. AW was supported by the Doctoral program ''Hadrons in Vacuum, Nuclei and Stars``, funded by the Austrian Science Fund FWF, contract W1203-N16.

\end{document}